\documentclass[aps,showpacs,preprintnumbers,amsmath, amssymb]{revtex4}

\usepackage{float}
\usepackage{graphics,epsfig}
\usepackage{graphicx}
\usepackage{dcolumn}
\usepackage{bm}

\begin{document}
\baselineskip=0.8 cm

\title{{\bf Holographic entanglement entropy in two-order insulator/superconductor transitions}}
\author{Yan Peng$^{1}$\footnote{yanpengphy@163.com},Guohua Liu$^{1}$}
\affiliation{\\$^{1}$ School of Mathematical Sciences, Qufu Normal University, Qufu, Shandong 273165, China}

\vspace*{0.2cm}
\begin{abstract}
\baselineskip=0.6 cm
\begin{center}
{\bf Abstract}
\end{center}

We study holographic superconductor model with two orders
in the five dimensional AdS soliton background away from the probe limit.
We disclose properties of phase transitions mostly from the holographic topological entanglement entropy approach.
Our results show that the entanglement entropy is useful in investigating
transitions in this general model and in particular, there is a new type of first order phase transition
in the insulator/superconductor system.
We also give some qualitative understanding and obtain
the analytical condition for this first order phase transition to occur.
As a summary, we draw the complete phase diagram representing effects of the scalar charge on phase transitions.

\end{abstract}

\maketitle
\newpage
\vspace*{0.2cm}

\section{Introduction}

The AdS/CFT correspondence provides us a novel approach to holographically study
strongly interacting system  in condensed matter physics where conventional calculational methods fail.
According to this correspondence,
the strongly interacting theories on the boundary are
dual to the higher dimensional weakly coupled gravity theories in the bulk \cite{Maldacena,S.S.Gubser-1,E.Witten}.
When applying a charged scalar field coupled to the Maxwell field in the bulk,
it was shown that the AdS black hole becomes unstable
to form scalar hair below a critical temperature, which is interpreted as a holographic metal/superconductor
transition on the boundary field theory \cite{S.A. Hartnoll,C.P. Herzog,G.T. Horowitz-1}.
Besides the metal/superconductor model, the holographic insulator/superconductor model
was also constructed in the AdS soliton spacetime \cite{TN,GTH}.
Such holographic superconductor models are interesting since they exhibit many
characteristic properties shared by real superconductor.
At present, a lot of more complete holographic superconductor models were
constructed, see Refs. \cite{R}-\cite{MB}.

It is known that the Ginzburg-Landau theory is
a powerful tool in understanding conventional superconductors.
The single-component Ginzburg-Landau theory was also generalized to
the two-component case and the two-component Ginzburg-Landau model was successfully
applied to the two-band systems \cite{MSE,AAM,AVA}.
On the holographic superconductor aspects, it is desirable to
study holographic models with multi-order parameter inspired by the two-component Ginzburg-Landau theory.
However,  most of holographic superconductor models existing in the literature involves only one order parameter.
In the background of AdS black hole, holographic metal/superconductor model
with two order parameters was firstly studied in the probe limit \cite{PBJ},
and then with back-reaction \cite{RCL,WW}.
Besides the s+s system, competition between $<p_{x}>$ and $<p_{x}+ip_{y}>$ orders was also studied in the generalized
p-wave model \cite{LAZ} and homogeneous/inhomogeneous phases competition appears in the presence of an interaction term \cite{JAE}. Moreover, superfluid, stripes and
metamagnetic phases may compete at low temperature when including large magnetic field in AdS black hole background \cite{ADP}.
On another side, due to the essential difference between the AdS black hole and AdS soliton,
the holographic insulator/superconductor transition with two scalar fields
was firstly studied in the four dimensional AdS soliton geometry in the probe limit \cite{RL}.
It was shown that there is a second order transition with coexist orders.
In contrast, we will construct a holographic $s+s$ model
in the five dimensional AdS soliton spacetime beyond the probe limit in this work.
Compared with results in \cite{RL},
we will show that our model allows new first order insulator/superconductor transitions that one order
condensation directly give way to another order condensation without coexist phases,
which has been observed in the metal/superconductor model \cite{ZN,MN}.

On the other side, the holographic entanglement entropy representing the degrees of freedom of
systems was recently applied to study properties of holographic phase transitions.
The authors provided an elegant approach to holographically
calculate the holographic entanglement entropy of
a strongly interacting system on the boundary from a
weakly coupled gravity dual in the bulk \cite{S-1,S-2}.
In this way, the holographic entanglement entropy has recently been
applied to disclose properties of phase transitions in various holographic
superconductor models \cite{NishiokaJHEP}-\cite{W}.
And the entanglement entropy turns out to be useful in
investigating the critical phase transition points and the order of
holographic phase transitions.
However, all the above discussions were carried out in the single order model.
In this paper, we initiate a discussion to examine whether the holographic entanglement entropy
approach is still valid in studying properties of holographic superconductor model with two order parameters.

The next sections are planed as follows. In section II,
we construct the two-order holographic insulator/superconductor model
in the five dimensional AdS soliton spacetime beyond the probe limit.
In section III, we observe various types of phase transitions by choosing different
values of the scalar charge. In particular, we manage to obtain a new first order phase transition
in holographic insulator/superconductor model.
We also draw complete phase diagram of the effects of the scalar charge on transitions.
We summarize our main results in the last section.

\section{Equations of motion and boundary conditions}

We are interested in the general holographic superconductor model dual to the bulk theory with two scalar fields
coupled to one single Maxwell field in the background of five dimensional AdS soliton.
And the corresponding Lagrange density in St$\ddot{u}$ckelberg form reads \cite{WW}:
\begin{eqnarray}\label{lagrange-1}
\mathcal{L}=R+\frac{12}{L^{2}}-\gamma[\frac{1}{4}F^{MN}F_{MN}-(\partial \psi_{1})^{2}-\psi_{1}^{2}(\partial \theta_{1}
-q A_{\mu})^{2}-m_{1}^{2}\psi_{1}^{2}-(\partial \psi_{2})^{2}-\psi_{2}^{2}(\partial \theta_{2}
-A_{\mu})^{2}-m_{2}^{2}\psi_{2}^{2}],
\end{eqnarray}

where $m_{1}$ and $m_{2}$ are the mass of the scalar fields $\psi_{1}(r)$ and $\psi_{2}(r)$ respectively.
$A_{M}$ stands for the ordinary Maxwell field.
 $-6/L^{2}$ is the negative cosmological constant with $L$ as the AdS radius.
In this work, we will adopt the convention that $L=1$ in the following calculation.
$\gamma$ represents the backreaction of matter fields on the background.
When $\gamma\rightarrow 0$, we go back to the holographic model in the probe limit.
We define $q_{1}, q_{2}$ as the charge of $\psi_{1}(r)$ and $\psi_{2}(r)$ respectively.
According to the lagrange density (1) , $q_{1}=q$ is the charge of the first scalar field $\psi_{1}(r)$ and without loosing generality, the charge of
the second scalar field $\psi_{2}(r)$ is set to $q_{2}=1$ under a transformation of the matter fields \cite{Y. Brihaye}.
In the following, we only consider the minimal model which gives the phase-locking condition,
saying $\theta_{1}=\theta_{2}=\theta$ \cite{EB}. Using the gauge symmetry
$A_{\mu}\rightarrow A_{\mu}+\partial \alpha, \theta\rightarrow \theta+\alpha$,
we can set $\theta=0$ without loss of generality.

Considering the form of the Lagrangian density and the matter fields' backreaction
on the metric, we take the deformed five dimensional AdS soliton solution as
\begin{eqnarray}\label{AdSBH}
ds^{2}&=&-r^{2}e^{C(r)}dt^{2}+\frac{dr^{2}}{r^{2}B(r)}+r^{2}dx^{2}+r^{2}dy^{2}+r^{2}B(r)e^{D(r)}d\chi^{2}.
\end{eqnarray}
Firstly, we need $C(r\rightarrow\infty)=0$ and $D(r\rightarrow\infty)=0$ to recover the asymptotic AdS boundary.
In order to get smooth solutions at the tip $r_{s}$ satisfying
$B(r_{s})=0$, we also have to impose on the coordinate $\chi$ a period
$\Gamma$ as
\begin{eqnarray}\label{HawkingT}
\Gamma=\frac{4\pi e^{^{-D(r_{s})/2}}}{r_{s}^{2}B'(r_{s})}.
\end{eqnarray}

For simplicity, we study matter fields with only radial dependence in the form

\begin{eqnarray}\label{symmetryBH}
A=\phi(r)dt,~~~~~~~~\psi_{1}=\psi_{1}(r),~~~~~~~\psi_{2}=\psi_{2}(r).
\end{eqnarray}

From above assumptions, we obtain equations of motion as
\begin{eqnarray}\label{BHpsi}
\psi_{1}''+\left(\frac{5}{r}+\frac{B'}{B}+\frac{C'}{2}+\frac{D'}{2}\right)\psi_{1}'+\frac{q^{2}\phi^{^{2}}
e^{-C}}{r^{4}B}\psi_{1}-\frac{m_{1}^{2}}{r^{2}B}\psi_{1}=0,
\end{eqnarray}
\begin{eqnarray}\label{BHpsi}
\psi_{2}''+\left(\frac{5}{r}+\frac{B'}{B}+\frac{C'}{2}+\frac{D'}{2}\right)\psi_{2}'+\frac{\phi^{^{2}}
e^{-C}}{r^{4}B}\psi_{2}-\frac{m_{2}^{2}}{r^{2}B}\psi_{2}=0,
\end{eqnarray}
\begin{eqnarray}\label{BHphi}
\phi''+\left(\frac{3}{r}+\frac{B'}{B}-\frac{C'}{2}+\frac{D'}{2}\right)\phi'-
\frac{2q^{2}\psi_{1}^{2}}{r^{2}B}\phi-\frac{2\psi_{2}^{2}}{r^{2}B}\phi=0,
\end{eqnarray}
\begin{eqnarray}\label{BHg}
C''+\frac{1}{2}C'^{2}+\left(\frac{5}{r}+\frac{B'}{B}+\frac{D'}{2}\right)C'-\gamma\frac{e^{-C}\phi'^{2}}{r^{2}}
-\gamma\frac{2q^{2}\phi^{2}\psi_{1}^{2}e^{-C}}{r^{4}B}-\gamma\frac{2\phi^{2}\psi_{2}^{2}e^{-C}}{r^{4}B}=0,
\end{eqnarray}
\begin{eqnarray}\label{BHChi}
B'\left(\frac{3}{r}-\frac{C'}{2}\right)+B\left(\gamma\psi_{1}'^{2}+\gamma\psi_{2}'^{2}-\frac{1}{2}C'D'
+\gamma\frac{e^{-C}\phi'^{2}}{2r^{2}}+\frac{12}{r^{2}}\right)\nonumber\\
+\gamma\frac{q^{2}\phi^{2}\psi_{1}^{2}e^{-C}}{r^{4}}+\gamma\frac{\phi^{2}\psi_{2}^{2}e^{-C}}{r^{4}}
+\gamma\frac{m_{1}^{2}\psi_{1}^{2}}{r^{2}}+\gamma\frac{m_{2}^{2}\psi_{2}^{2}}{r^{2}}-\frac{12}{r^{2}}=0,
\end{eqnarray}
\begin{eqnarray}\label{BHChi}
D'=\frac{2r^{2}C''+r^{2}C'^{2}+4rC'-2\gamma e^{-C}\phi'^{2}+4\gamma r^{2}\psi_{1}'^{2}+4\gamma r^{2}\psi_{2}'^{2}}{r(6+r C')}.
\end{eqnarray}
These equations are nonlinear and coupled, so we have to use the
shooting method to search for the numerical solutions satisfying boundary conditions.
At the tip, the solutions can be expanded as
\begin{eqnarray}\label{InfBH}
&&\psi_{1}(r)=\psi_{10}+\psi_{11}(r-r_{s})+\cdots,~~~~\psi_{2}(r)=\psi_{20}+\psi_{21}(r-r_{s})+\cdots,\nonumber\\
&&\phi(r)=\phi_{0}+\phi_{1}(r-r_{s})+\cdots,~~~~~~~~B(r)=B_{0}(r-r_{s})+\cdots,\nonumber\\
&&C(r)=C_{0}+C_{1}(r-r_{s})+\cdots,~~~~~~D(r)=D_{0}+D_{1}(r-r_{s})+\cdots,
\end{eqnarray}
where the dots denote higher order terms.
Putting these expansions into equations of motion and considering the leading
term, we have six independent parameters $r_{s}$,
$\psi_{10}$, $\psi_{20}$, $\phi_{0}$, $C_{0}$ and $D_{0}$ left to describe the solution.
Near the infinity boundary $(r\rightarrow \infty)$, the asymptotic
behaviors of the solutions are
\begin{eqnarray}\label{InfBH}
&&\psi_{i}\rightarrow\frac{\psi_{i-}}{r^{\lambda_{i-}}}+\frac{\psi_{i+}}{r^{\lambda_{i+}}}+\cdots,~~~
\phi\rightarrow \mu-\frac{\rho}{r^{2}}+\cdots,~~~ \nonumber\\
&&B\rightarrow 1+\frac{B_{4}}{r^{4}}+\cdots,~~~ C\rightarrow
\frac{C_{4}}{r^{4}}+\cdots,~~~ D\rightarrow
\frac{D_{4}}{r^{4}}+\cdots,
\end{eqnarray}

with $\lambda_{i\pm}=(2\pm\sqrt{4+m_{i}^{2}})$.
The coefficients above can be related to physical quantities in the boundary field theory
according to AdS/CFT dictionary. $\mu$ and $\rho$ can be
interpreted as the chemical potential and the density of the charge carrier in the dual
theory respectively. $\psi_{i+}=<O_{i+}>$ are the operator parameters
in the CFT on the boundary, where $i=1,2$. The parameters $B_{4}$, $C_{4}$ and $D_{4}$
are integration constants.

The equations of motion are invariant under the scaling symmetry
\begin{eqnarray}\label{symmetryBH}
r \rightarrow ar,~~~~~~~~(\chi,x,y,t)\rightarrow~(\chi,x,y,t)/a,~~~~~~~\phi\rightarrow
a\phi,
\end{eqnarray}
which can be used to set $r_{s}=1$ in the numerical calculation.
Choosing $m_{1}^{2}=0$ and $m_{2}^{2}=\sqrt{-\frac{15}{4}}$ above the BF bound $m_{BF}^{2}=-4$
\cite{P. Breitenlohner}, the modes $\psi_{i+}$ are always normalizable
and can be interpreted as the expectation value of operators in a dual theory
with dimension $\lambda_{i +}$.
To get a stable theory, we will fix the first operators $\psi_{i-}=0$ and use the second
operators $\psi_{i+}=<O_{i+}>$ to describe the phase transition in the dual CFT.

\section{Properties of scalar condensation}

In this part, we concentrate on the holographic entanglement entropy(HEE) of the transition system.
The authors in Refs. \cite{S-1,S-2} have proposed a novel way to calculate
the entanglement entropy of strongly interacting systems in conformal field theories (CFTs)
on the boundary from a weakly coupled AdS spacetime in the bulk.
For simplicity, we choose to study the entanglement
entropy for a half space with the subsystem $\bar{A}$ defined as $x > 0$,
$-\frac{R}{2}<y<\frac{R}{2}$ ($R \rightarrow \infty$), $0 \leqslant \chi \leqslant \Gamma$.
Then the entanglement entropy can be expressed with the metric solutions in the form \cite{RC1,RC2,W}:

\begin{eqnarray}\label{InfBH}
S_{\bar{A}}^{half}=\frac{R\Gamma}{4G_{N}}\int_{r_{0}}^{\frac{1}{\varepsilon}}re^{\frac{D(r)}{2}}dr=\frac{R\pi}{8G_{N}}(\frac{1}{\varepsilon^{2}}+S_{EE}),
\end{eqnarray}
where $r=\frac{1}{\varepsilon}$ is defined as the UV cutoff. The first term depending on the UV cutoff
is divergent as $\varepsilon \rightarrow 0$.
In contrast, the second term is finite and independent of the UV cutoff.
So the second finite term is physical important.
In fact, this finite term is the difference of the entanglement entropy between the deformed
AdS soliton and the pure AdS space.
In the case of normal AdS soliton, the entanglement entropy is a constant: $S_{EE}=-1$.

It is known that the physical procedure is along the lowest free energy.
In Fig. 1, we plot the free energy of the case with $\gamma=0.1$, $\Gamma=\pi$,
$m_{1}^{2}=0$, $m_{2}^{2}=-\frac{15}{4}$ and $q=1$.
We find two possible phases corresponding to the condensation of different scalar fields.
For every fixed value of the chemical potential, we can choose only one phase.
It can be seen from fig. 1 that the solid blue line with the lowest free energy and a critical chemical potential $\mu=1.888$
is physical. It means the operators compete and the condensation of $<O_{2+}>$ prevents
another operator $<O_{1+}>$ to condense.
It was mentioned in \cite{PBJ} that this holographic property is similar to behaviors of real
weakly interacting Fermi liquids, where the condensation of one order tends to produce
a mass gap at least for parts of the Fermi surface and inhibits any other orders to condense.
We have to point out that we
only consider electric field and investigate homogeneous solutions in this work.
If including magnetic field, there may be inhomogeneous instabilities intervene
before the homogeneous condensation. For example, in the presence of magnetic
fields and an interaction term, it was shown in \cite{JAE} that spatially modulated
phases may have a larger critical temperature compared to the homogeneous
solution in AdS black hole background. And we plan to examine the influence
of the magnetic field on the phase structure in the next work.
\begin{figure}[h]
\includegraphics[width=230pt]{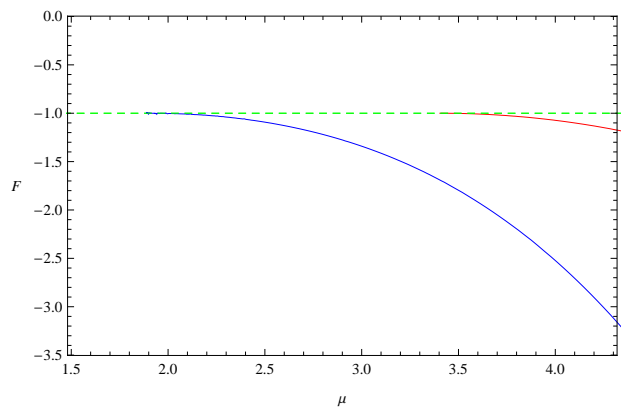}\
\caption{\label{EEntropySoliton} (Color online) The free energy in cases of
$\gamma=0.1$, $\Gamma=\pi$,$m_{1}^{2}=0$, $m_{2}^{2}=-\frac{15}{4}$ and  and $q=1$. The red line shows the condensation with
$<O_{1+}>\neq0$ and $<O_{2+}>=0$.
The blue curve represents the condensation $<O_{1+}>=0$ and $<O_{2+}>\neq0$. And the dashed green line is the free energy of the normal AdS soliton solution.}
\end{figure}

Now we study the case of $\gamma=0.1$, $\Gamma=\pi$,
$m_{1}^{2}=0$, $m_{2}^{2}=-\frac{15}{4}$ and $q=1$
through the behaviors of holographic entanglement entropy in Fig. 2.
The jump of the slop of the holographic entanglement entropy
with respect to the chemical potential around the threshold value $\mu_{c}=1.888$ is a sign of the second order phase transition
since this critical chemical potential is equal to the critical chemical potential $\mu=1.888$ obtained from behaviors of the free energy.
These properties of the holographic entanglement entropy is the same as the single scalar field case.
It is natural since our model with two scalar fields can be reduced to the holographic model
with only one scalar field when one scalar field condensation prevents another scalar field to condense.
\begin{figure}[h]
\includegraphics[width=230pt]{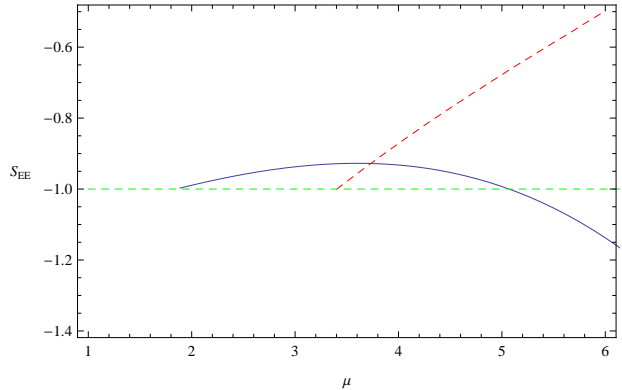}\
\caption{\label{EEntropySoliton} (Color online) The holographic entanglement entropy of the system of
$\gamma=0.1$, $\Gamma=\pi$, $m_{1}^{2}=0$, $m_{2}^{2}=-\frac{15}{4}$ and $q=1$. The dashed red line shows the
unstable phases $<O_{1+}>\neq0$ and $<O_{2+}>=0$.
The solid blue curve represents the physical phase of  $<O_{1+}>=0$ and $<O_{2+}>\neq0$.
And the dashed green curve corresponds to the normal AdS soliton solution.}
\end{figure}

We also manage to obtain complicated solutions
with another set of parameters
as: $\gamma=0.1$, $\Gamma=\pi$,
$m_{1}^{2}=0$, $m_{2}^{2}=-\frac{15}{4}$ and $q=2$.
We show the free energy of the system in Fig. 3.
We mention that the solid green line corresponds to phases with two nonzero orders.
However, it can be seen from behaviors of the free
energy that these phases are thermodynamically unstable.
As we increase chemical potential along the lowest free energy, the scalar field $\psi_{1}(r)$ firstly condenses at $\mu_{1}=1.702$
and then it give way to the condensation of the second scalar field $\psi_{2}(r)$ around $\mu_{2}=2.797$.
Moreover, the phase transition at $\mu_{1}$ is of the second order and
there is a first order phase transition around $\mu_{2}$, which is very different from former
related results in \cite{RL}.

\begin{figure}[h]
\includegraphics[width=200pt]{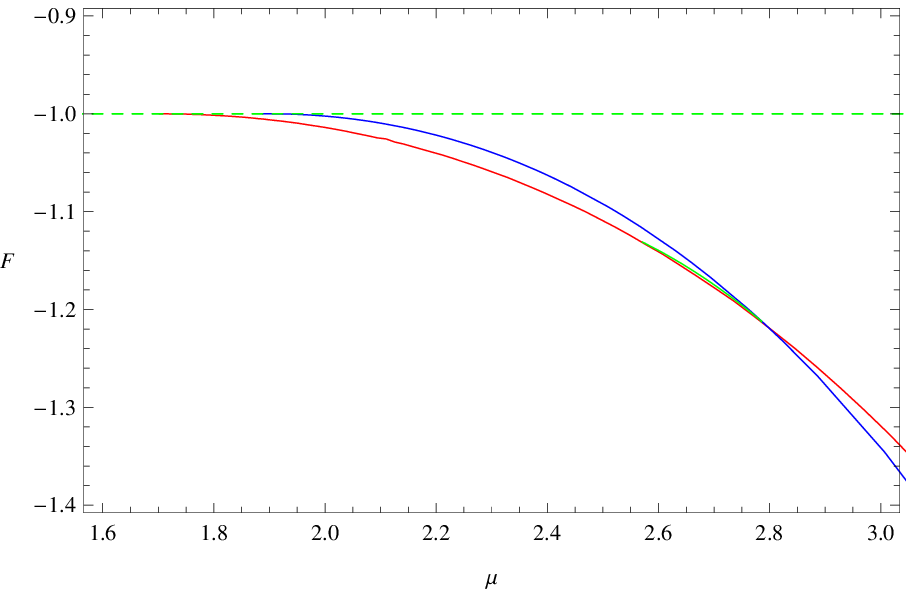}\
\includegraphics[width=200pt]{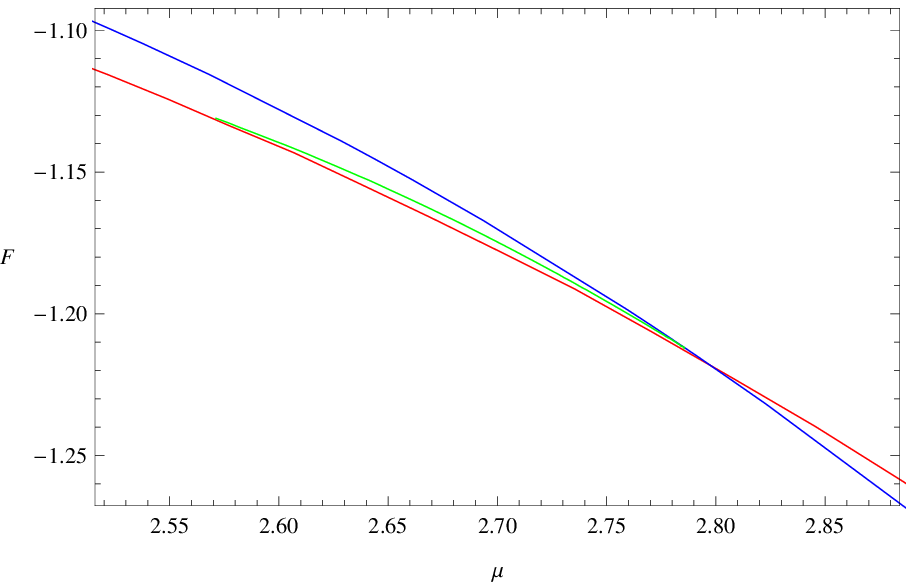}\
\caption{\label{EEntropySoliton} (Color online) In the left panel, we plot the free energy as a function of the chemical potential
with $\gamma=0.1$, $\Gamma=\pi$, $m_{1}^{2}=0$, $m_{2}^{2}=-\frac{15}{4}$ and $q=2$.
The solid red line is for the case of $<O_{1+}>\neq0$ and $<O_{2+}>=0$. The solid blue line is
with $<O_{1+}>=0$ and $<O_{2+}>\neq0$.
And the solid green curves in the range of $\mu\in[2.57,2.80]$ represent the coexisting phases of $<O_{1+}>\neq0$ and $<O_{2+}>\neq0$.
The dashed green curve corresponds to the normal phase.
In the right panel, we show a enlarged view in the coexisting phases region, which suggests
the coexisting phases are unstable.}
\end{figure}

Now we turn to study the transition through the holographic entanglement entropy method.
We show the holographic entanglement entropy with respect to the chemical potential
$\mu$ in Fig. 4 in cases of $\gamma=0.1$, $\Gamma=\pi$ and $q=2$.
The red line is the case of $m^{2}=0$
and the blue curve shows the case of $m^{2}=-\frac{15}{4}$.
At the second order phase transition points $\mu_{1}=1.702$,
there is a jump of the slop of the entanglement entropy with respect to the chemical potential.
It also can be easily seen from the panel that
the entanglement entropy itself also has a jump
at the first order phase transition points $\mu_{2}=2.797$.
It means there is a reduction in the number of degrees
of freedom due to the formation of new condensation.
In summary, the holographic entanglement entropy
approach is still useful in searching for the second order phase transition points
and also the order of phase transitions in this two-order holographic superconductor model.
\begin{figure}[h]
\includegraphics[width=200pt]{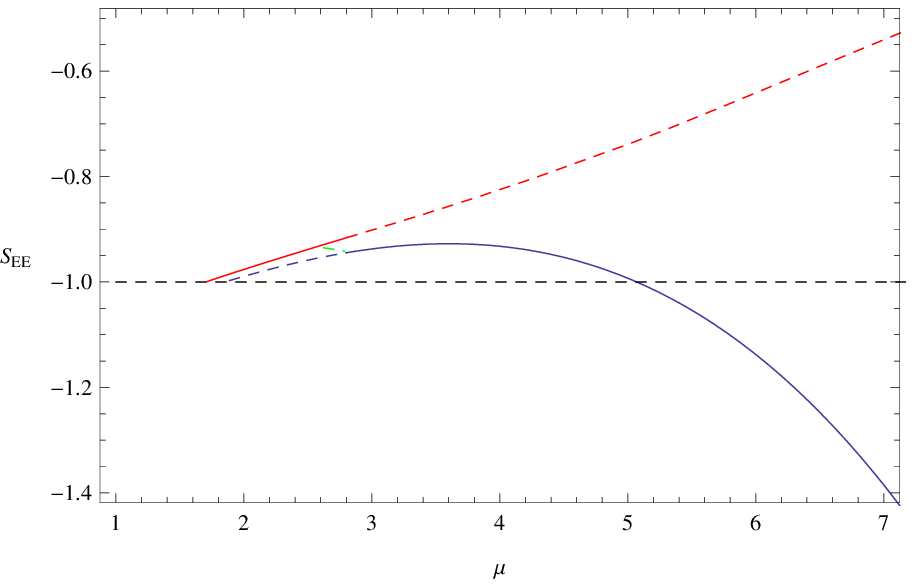}\
\includegraphics[width=200pt]{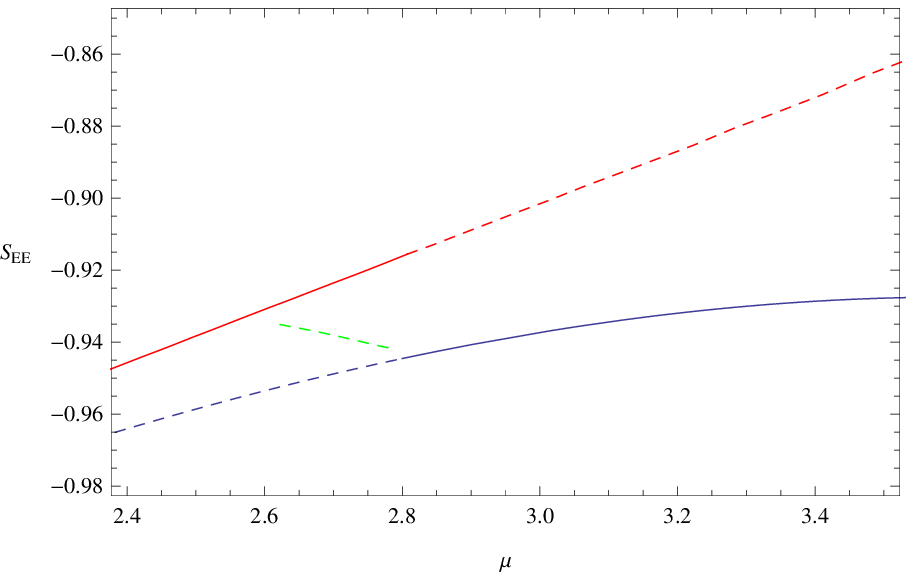}\
\caption{\label{EEntropySoliton} (Color online) We plot the holographic entanglement entropy with
$\gamma=0.1$, $\Gamma=\pi$, $m_{1}^{2}=0$, $m_{2}^{2}=-\frac{15}{4}$ and $q=2$ in the left panel. The red line is with $<O_{1+}>\neq0$ and $<O_{2+}>=0$,
the blue curve represents the case of $<O_{1+}>=0$ and $<O_{2+}>\neq0$
and the dashed green line corresponds to coexist phases.
We plot the pure AdS soliton state with dashed black line.
The solid lines correspond to physical superconducting phases.
In the right panel, we show a enlarged view in the coexisting phase region. Here the solid lines correspond
to stable phases.}
\end{figure}

By studying in detail behaviors of the holographic entanglement entropy,
we go on to disclose the complete phase structure
that how the scalar field charge could affect the critical chemical potential
and the order of phase transitions in Fig. 5 with $\gamma=0.1$, $\Gamma=\pi$, $m_{1}^{2}=0$ and $m_{2}^{2}=-\frac{15}{4}$.
It can be easily seen from the picture that there is a critical charge $q_{c}=1.81$,
above which one operator $<O_{1+}>$ firstly condenses through a second order phase transition and then
it transforms into the condensation of another operator $<O_{2+}>$
through a first order phase transition. In contrast,  for $q<1.81$, the normal phase directly
changes into phases with $<O_{2+}>\neq0$ and the operator $<O_{1+}>$ will never condenses
when we increase the chemical potential.

\begin{figure}[h]
\includegraphics[width=230pt]{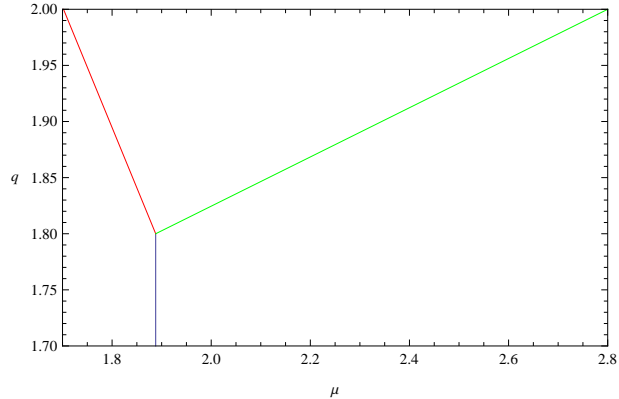}\
\caption{\label{EEntropySoliton} (Color online) We show the effects of the charge on the critical chemical potential and
the order of phase transitions with $\gamma=0.1$, $\Gamma=\pi$, $m_{1}^{2}=0$ and $m_{2}^{2}=-\frac{15}{4}$. The left area is the pure AdS soliton with $<O_{1+}>=0$ and $<O_{2+}>=0$,
the top area corresponds to the condensation with $<O_{1+}>\neq0$ and $<O_{2+}>=0$ and the right area represents the condensation
with $<O_{1+}>=0$ and $<O_{2+}>\neq0$. The red line and the blue line are second order phase transitions. The green curve
between different order condensation corresponds to the first order phase transitions.}
\end{figure}

At last, we give some analytical understanding of the properties obtained from numerical solutions.
When constructing the first order phase transition,
we have applied two scalar fields with different critical chemical potential
and assume the scalar field with larger critical chemical potential has a more negative mass.
In this way, the scalar field with a smaller critical potential firstly condenses and then
another scalar field with a more negative scalar mass quickly plays a dominant
role in the condensation. This mechanism produces the phase structure that
one order parameter condensation transforms into another order parameter condensation.
We can easily generalize the approximate formula for critical chemical potential in \cite{Yan}
to the form: $\mu_{c}\thickapprox\frac{a+\sqrt{m^{2}-m_{BF}^{2}}}{q}$ with $a=1.400$.
Based on this approximate formula,
we give an analytical condition that the first order phase transitions could happen as:
$\frac{a+\sqrt{m_{1}^{2}-m_{BF}^{2}}}{q}<a+\sqrt{m_{2}^{2}-m_{BF}^{2}}$ with $m_{1}^{2}>m_{2}^{2}$ and $a=1.400$.
In Fig. 5, we have $m_{1}^{2}=0$, $m_{2}^{2}=-\frac{15}{4}$ and $m_{BF}^{2}=-4$.
Then we have a threshold value $q_{c}=1.80$, above which there is first order phase transitions.
This analytical result $q_{c}=1.80$ is in good agreement with the numerical result $q_{c}=1.81$.

\section{Conclusions}

We explored properties of holographic insulator/superconductor transitions with two orders
in the five dimensional AdS soliton background beyond the probe limit.
We investigated phase transitions mostly
from the holographic topological entanglement entropy approach.
We showed that the holographic topological entanglement entropy
is useful in determining the second order critical chemical potential and the
order of phase transitions in this two-order model.
For different sets of parameters, we observed various types of transitions.
In particular, we found transitions with properties
that one order parameter condensation is transformed into another,
which is a new type of first order phase transition in the insulator/superconductor model.
We also gave some qualitative understanding and obtain
the precise analytical condition for the first order phase transition to occur.
As a summary, we obtained the complete phase diagram from behaviors of the
holographic topological entanglement entropy.

\begin{acknowledgments}

This work was supported
by the National Natural Science Foundation of China under Grant No. 11305097;
the Shaanxi Province Science and Technology Department Foundation of China
under Grant Nos. 2016JQ1039 and 2016JM1028.

\end{acknowledgments}

\end{document}